\documentclass[aps,prd,preprint,tightenlines,groupedaddress,showpacs]{revtex4-1}
\usepackage{epsf,epsfig,graphics,graphicx}
\usepackage{dcolumn}
\usepackage{bm}

\begin{document}


\title
{Constraining the dark-energy equation of state with cosmological data}

\author{Yu-Ping Teng$^1$}
\author{Wolung Lee$^1$}
\author{Kin-Wang Ng$^{2,3}$}

\affiliation{
$^1$Department of Physics, National Taiwan Normal University,
Taipei 11677, Taiwan, Republic of China\\
$^2$Institute of Physics, Academia Sinica, Taipei 11529, Taiwan, Republic of China\\
$^3$Institute of Astronomy and Astrophysics, Academia Sinica, Taipei 11529, Taiwan, Republic of China}

\date{\today}

\begin{abstract}
Recently, the observed equation of state for dark energy appears to favor values below $-1$. The tendency implies that the nature of dark energy may be quite different from that of the cosmological constant.  In view of the adjustment on the equation of state keeps decreasing, the introduction of the phantom energy seems inevitable. By employing observational constraints from supernovae and from the acoustic scale in which the accuracy of the data has become extraordinary, we apply a phenomenological scenario to be acquainted with the evolution of our universe. The demonstration on the constrained unfolding of the phantom energy shows the model has high consistency with the current observation. 
\end{abstract}

\maketitle

\section{\label{sec:level1}Introduction}

The measurements in cosmological parameters have developed in full swing in recent years. Measurements including the Planck collaboration, the Wilkinson Microwave Anisotropy Probe (WMAP), the Sloan Digital Sky Survey (SDSS) on baryon acoustic oscillations (BAO), and many local distance determinations on supernovae have become astonishingly precise and are constrained below percent level. Predictions from the standard $\Lambda$CDM model on certain key parameters such as the Hubble constant $H_0$, the space-time geometry, and the sound horizon $r_s$, significantly facilitate our understanding to the evolution of the universe. The indirect, or to say, the model dependent measurement derived from the data set of Planck collaboration 2018~\cite{01aghanim2018planck} renders accurate numbers on $H_0$ and $r_s$ which serve as the absolute scale for the distance measurement at the opposite epochs of the cosmos ~\cite{02cuesta2015calibrating,03aubourg2015cosmological}. Usually, the determination of cosmic distances is closely related to the assumption of the fundamental model describing the evolutionary process of the universe. For instance, the present expansion rate, $H_0$, which is constrained by and can be derived from angular anisotropies in the cosmic microwave background (CMB) may be disparate if the underlying theory has a nuance. Nevertheless, the sound horizon at the radiation drag epoch, $r_s$, seems to only rely upon the physics at the early times where it is not that controversial in various models. Moreover, the BAO provides a consistency check on the evolving speed in late eras. Consequently, the large scale structure emerged from the primordial matter perturbation offers a standard ruler for the distance measurement. Therefore, one is able to pin down and retrieve the Hubble parameter $H(z)$ in cosmic times. 

Due to the model dependency of the Planck collaboration on measuring $H_0$, the local distance measurement is contrived to offer an independent check for the sake of consistency~\cite{camarena2020local}. The Planck collaboration 2018~\cite{01aghanim2018planck} shows that the Hubble constant should be constrained at $H_0 = (67.27\pm0.60)$ km/s/Mpc. However, the recent measurement on SNeIa calibrated by Cepheid variables~\cite{04riess2018new} promotes that number to $H_0 = (73.48\pm1.66)$ km/s/Mpc, thus creating a tension to Planck's outcome up to 3.7$\sigma$. In fact, SN data in recent years yields similar results. For examples, $H_0 = 74.22\pm1.82$ km/s/Mpc in~\cite{riess2019large} and $ 73.2\pm1.3$ km/s/Mpc in~\cite{riess2021cosmic}. On the other hand, the project ``$H_0$ Lenses in Cosmograil's Wellspring" (H0LiCOW) measuring the time-delay cosmography of quasars by strong lensing provides a model-independent expansion rate as $H_0 = 71.9^{+2.4}_{-3.0}$ km/s/Mpc for a flat universe with free matter and energy densities~\cite{05bonvin2017h0licow}. That will be upgraded to $H_0 = (72.8\pm2.4)$ km/s/Mpc if taking the matter density $\Omega_m = 0.32$ into account as observed by the Planck collaboration. The H0LiCOW measurement has been further updated by TDCOSMO analysis which renders $H_0 = 74.5^{+5.6}_{-6.1}$~\cite{birrer2020tdcosmo}. On the other hand, the red giant branch (TRGB) calibration~\cite{freedman2019carnegie} offers another model-independent measurement as $H_0 = 69.8\pm1.7$ km/s/Mpc, which is still $1.2\sigma$ away from the Planck's result. The differences in all prime measurements can be easily seen in Fig. 1 of~\cite{di2021realm}. Such a tension between model dependent and independent measurements seems to be an axiomatic problem. As the accuracy increasing on each measurement the tension, however, is exacerbated. 

In order to alleviate this tension, the physics beyond the standard model has been provided including higher number of effective relativistic species, extended cosmological parameters~\cite{06di2016reconciling,di2020cosmological}, and the dynamically evolving dark energy~\cite{07aubourg2015cosmological,08ade2016planck,dde2020,yang2021dynamical}. In particular, scenarios with a dynamical dark energy do have the advantage of describing the mechanism of the cosmic acceleration~\cite{10di2017constraining}. The value of the dark-energy equation of state obtained by the Nine-Year Wilkinson Microwave Anisotropy Probe (WMAP9)~\cite{11hinshaw2013nine} which combined data from the CMB, BAO, supernovae, and $H_0$ measurements indicates that $w = -1.084 \pm 0.063$. The Planck collaboration reported in the early 2015 that $w = -1.006 \pm 0.0451$~\cite{12ade2016planck}, and in 2018, $w = -1.028\pm0.032$~\cite{01aghanim2018planck} (68 $\%$, Planck TT,TE,EE+lowE+lensing+SNe+ BAO). Since the dark-energy equation of state provided by the Planck has reached the edge of $-1$, the inclusion of a phantom component (i.e. a dark energy with  an equation of state less then $-1$) seems inevitable. Furthermore, once allowing the equation of state evolving as $w(z) = w_0+(1-a)w_a$, where $a$ denotes the cosmic scale factor, the constraint derived from CMB+lensing+BAO~\cite{01aghanim2018planck} conveys a message that a universe containing a phantom component is more likely to happen than the one that carries a quintessence dark energy with $w > -1$~\cite{banerjee2021hubble}. In fact, it is arguably that the very existence of a phantom energy seems unavoidable given the predicament of the $H_0$ tension~\cite{vagnozzi2020,vagnozzi2018}. 

We consider the evolution of the universe with a dynamically evolving dark energy in this paper. The energy density of a phantom like component must be extremely low in the remote past owing to its lopsided negative equation of state. As a consequence, parameterizing the dynamical dark energy is rather delicate ~\cite{chevallier2001accelerating,linder2003exploring,cooray1999gravitational,gong2005probing,yang2019observational} and can be highly biased~\cite{colgain2021can}. However, utilizing the absolute scale (surveyed by model-independent measurements) in two contrary epochs, the present and the last scattering surface, the consistency in parameters can be checked by tracing the cosmic evolution through the dark age facilitated by various models of dynamical dark energy. Currently, there are several theories regarding the phantom energy stipulating its own construction upon the potential field to support such an exotic existence, e.g. the  vacuum phase transition ~\cite{13parker2000new,14parker2004acceleration,15caldwell2006sudden,khosravi2019h,banihashemi2020phase}, Dvali-Gabadadze-Porrati branes~\cite{16koyama2007ghosts,20dvali20004d}, vector fields~\cite{17armendariz2004could,18jimenez2008cosmic,19koivisto2008vector}, interacting dark energies~\cite{kumar2016probing,di2017can}, kinetic braidings~\cite{21deffayet2010imperfect}, and other scalar tensor varieties~\cite{23bluhm2005spontaneous,24elder2014satisfying}. Though all the hypotheses could be pragmatic for reconciling the tension between the cosmological parameters, we avoid such sophisticated assumptions in our calculation by adopting a relatively simple formula to construct the phantom component. We present our numerical scheme in Sec. II, and the results of the calculation in Sec. III. Finally, we discuss about the requisite of the phantom energy under current releasing data of all measurement in Sec. IV. 

\section{\label{sec:level2}Method}  

Despite lacking sufficient clue, a generalized phenomenological approach called the generic quintessence (GQ)~\cite{25GQ} is very helpful in exploring the dark energy with limited observational figures. Yet while identifying the potential field for driving the late time cosmic acceleration remains an important subject, the numerical reconstruction of the dark-energy equation of state is an efficacious method to retrieving useful information about the nature of dark energy. Instead of presupposing the Lagrangian for dark energy and deriving the equation of state from a potential field, we assume a rather simple form for the equation of state while allowing for proper adjustments in the course of evolution. The necessity of a phantom component is much easier to examine under this scheme. The Hubble constant $H_0$ and the sound horizon $r_s$, which play as characteristic scales on the contrary side of evolution, will be the fixation of our calculation. Thereupon the model-independent measurements on these two parameters are the legitimate choices. 

Consider a flat universe where the total density parameter $\Omega_0$ is characterized by $\Omega_0 = \Omega_{m,0}+\Omega_{r,0}+\Omega_{\phi,0} = 1$ with the matter density $\Omega_{m,0}\sim0.3$, the density of the dark energy $\Omega_{\phi,0}\sim0.7$, and a negligible radiation density $\Omega_{r,0}$~\cite{12ade2016planck}. The background evolution of the $i$th component is governed by
\begin{eqnarray}
&\dot{\varepsilon_i}+3H(1+w_i)\varepsilon_i = 0,
\label{eq:one}
\\
&\dot{H}+\frac32H^2+\sum_i\frac{w_i}{2M^2_p}\varepsilon_i = 0,
\label{eq:two}
\end{eqnarray}
where $i=r$, $m$ and $\phi$ respectively. In terms of the reduce Planck mass $M_p=(8\pi G)^{-1/2}$, the conformal time interval $d\eta = H_0a^{-1}dt$, the dimensionless Hubble constant $\mathcal{H} = H/H_0$, and the rescaled energy density $\rho_i = \varepsilon_i/(M_pH_0)^2$, the evolution can be recast as
\begin{eqnarray}
\frac{d\rho_i}{d\eta}& = &-3a\mathcal{H}(1+w_i)\rho_i
\label{eq:three},
\\
\frac{d\mathcal{H}}{d\eta}& = &-\frac32a\mathcal{H}^2-\sum_i\frac{w_i}{2}a\rho_i
\label{eq:four},
\\
\frac{da}{d\eta}& = &a^2\mathcal{H}
\label{eq:five},
\\
\frac{d\tau}{d\eta}& = &a
\label{eq:six},
\end{eqnarray}
where the numerical equation of state $w_i(\eta)$ are designated for tracing the underlying model intimately.  In particular, we use an $\Omega_\phi$-weighted average to constrain the dynamic behavior of the dark-energy component, i.e. 
\begin{eqnarray}
\left<w_\phi\right>= \int_{\eta_{\star}}^{\eta_0}\Omega_\phi(\eta)w_\phi(\eta)d\eta/\int_{\eta_{\star}}^{\eta_0}\Omega_\phi(\eta)d\eta
\label{eq:seven},
\end{eqnarray}
where $\eta_0$ and $\eta_\star$ represent the conformal times at the present and at the last scattering surface respectively. As far as the dark-energy component is not efficient to significantly changing the CMB anisotropy spectrum, the approximation using a constant $\left<w_\phi\right>$ is effective for models to be tested~\cite{26bean2002current,27baccigalupi2002constraints,28corasaniti2002constraining,29huey1999resolving}.

\begin{figure}
	\scalebox{0.7}
	{\includegraphics{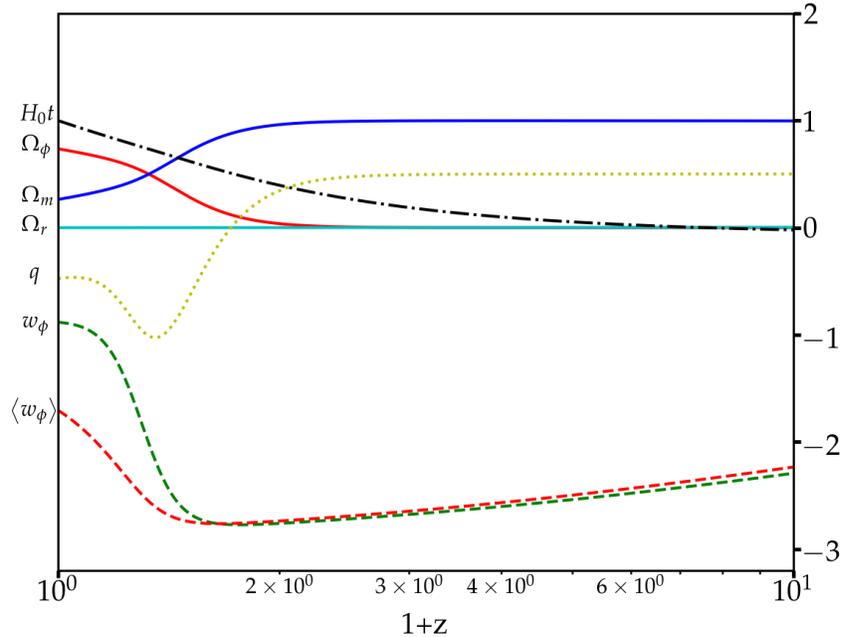}}
	\caption{\label{fig:example} An example of the cosmic background evolution with a dynamical dark energy component maintaining $w_\phi(z)<-1$.}
\end{figure}

Applying the initial conditions at the present time, the problem becomes a drill of solving a set of first-order ordinary differential equations. Since each relevant parameter may vary drastically, we employ the ODE solving function 'dopri5' with adaptive step size to solve for the rapidly evolving stiff system. As an example, Fig.~\ref{fig:example} illustrates a toy universe with a phantom component possessing $w_\phi(z)<-1$ against the cosmological redshift $z$.
For later comparisons and discussions, we now specify all relevant parameters in our calculation. The initial conditions are provided by the current observational data. In the energy inventory~\cite{01aghanim2018planck}, we have $\Omega_{b,0}h^2 = 0.02242\pm0.00014$ for baryons, $\Omega_{m,0}h^2 = 0.1424\pm0.00087$ for all matters, and $\Omega_{\phi,0}=\Omega_\Lambda = 0.6889\pm0.0056$ for the unknown dark energy at the present time. On the other hand, the Hubble constant may change significantly from models to models. Under the circumstances, we take $H_0 = (73.48\pm1.66)$ km/s/Mpc from the local measurement on the type Ia supernova~\cite{04riess2018new} to check out its consistency. 

The sound horizon $r_s$, the other key parameter in our calculation, is described by the relation that
\begin{equation}
r_s = \int_{0}^{\eta_\star}c_sd\eta,
\end{equation}
in which the sound speed $c_s$ is determined by $c_s = 1/(3+R)$ with $R$ approximated as~\cite{30skordis2002planck} $30230(\Omega_bh^2)(1+z)^{-1}$. However, it can be linked to the acoustic scale $l_A$ derived from the Doppler peaks in the CMB angular power spectrum. According to the baseline model~\cite{31pan2016cosmic}, the anisotropy amplitude is spiking at $l_A = p\pi/\theta_\star = 302p$, in which the acoustic scale is defined as 
\begin{eqnarray}
l_A = {\pi\over \theta_\star}\equiv {\pi D_A\over r_s}
\label{eq:eight},
\end{eqnarray}
where $D_A = \eta_0-\eta_\star$ denotes the comoving distance to the last scattering surface. Subsequently, the sound horizon at the decoupling epoch with $z_\star = 1089$ must have subtended an angle as~\cite{01aghanim2018planck} 
\begin{eqnarray}
\theta_\star = \frac{r_s}{\eta_0-\eta_\star} = (1.04109\pm0.0003)\times10^{-2} 
\label{eq:nine}.
\end{eqnarray}
The error of the measurement is down to 0.03\% level. With such an extraordinary precision and the simple geometrical interpretation, the angular size of the sound horizon $\theta_\star$ becomes a model-independent baseline to examining the consistency of the cosmic evolution deep into the matter dominant phase. 

The calculation only bases upon the physical process prior to the last scattering surface and the model-independent observational result on the CMB anisotropies. Both are relatively insensitive to the underlying dark-energy component. Accordingly, the precision of the angular size, $\Delta\theta_\star = (\theta_{\star,cal}-\theta_{\star,obs})/\theta_{\star,obs}$, will lead us to the superior model under consideration. Since we evade the detailed behavior of the dark-energy component, we set the equation of state into specific values in different epochs. The maneuver actually helps us on interpreting the effect brought about by the equation of state beneath $-1$. Meanwhile, the comoving distance $D_A$ requires appropriate estimations on $\eta_\star$ and $\eta_0$. The approximation we used to handle this issue is described in Appendix~\ref{app:etaapp} where the conformal times are evaluated by substituting $\Omega_\phi$ with the dark energy density around the last scattering surface $\Omega_{\phi,\star}$, and the equation of state $w_\phi$ with the time-average $\left<w_\phi\right>$.

The angular diameter distance $D_A$ and the Hubble parameter $H(z)$ can be determined by virtue of the measurement on the large scale structure in different redshifts. As the dynamical dark energy is capable of fitting the constraint in opposite eras, we need another clue in between to trace out the whole picture of evolution. Thus we set the equation of state as a justifiable free parameter in a flexible fashion to fit the recent BAO measurements. We employ the data that $H(z_{\rm eff}=0.38,0.51,0.61) = 81.5\pm1.9, 90.4\pm1.9, 97.3\pm2.1$ from the SDSS-III Baryon Oscillation Spectroscopic Survey (BOSS DR12)~\cite{32alam2017clustering}, $H(z_{\rm eff}=1.52) = 162\pm12.0$ from the SDSS-IV DR14 quasar sample~\cite{33gil2018clustering}, and $H(z_{\rm eff}=2.36)=226\pm8.0$ from the Quasar-Lyman $\alpha$ from BOSS DR11 (BOSS Ly-$\alpha$)~\cite{34font2014quasar} to carry out the consistency check. 

\section{\label{sec:level3}Result}

\begin{figure}
	\scalebox{0.7}
	{\includegraphics{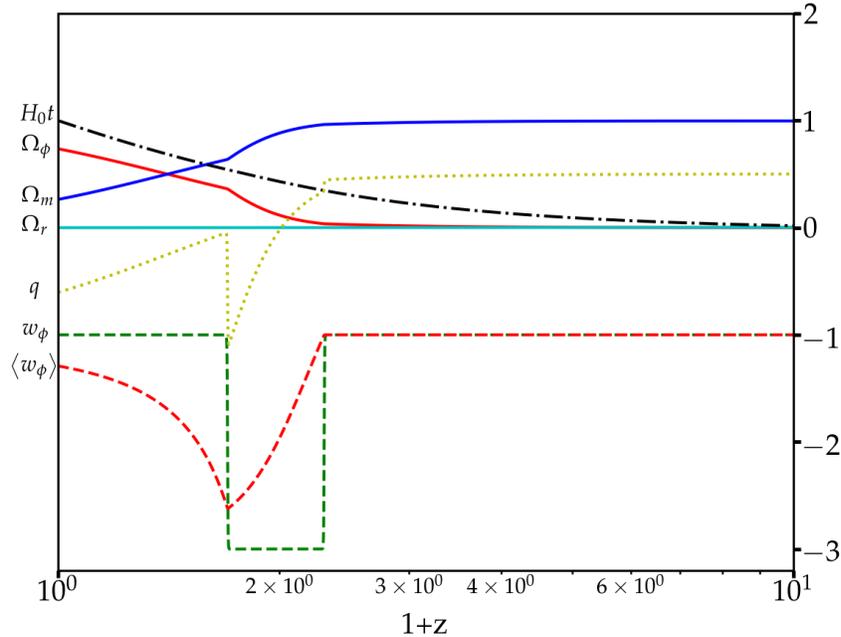}}
	\caption{\label{fig:phantom} The cosmic evolution under the vacuum metamorphosis model in the context of GQ's numerical scheme. Note the time-average dark-energy equation of state is signified by $\left<w\right>=-1.29$, and the error on $\theta_\star$ is $0.049\%$ in this example. }
\end{figure}

The constraints from the Planck collaboration and the local distance measurement have revealed a preference for the dark-energy component with $w<-1$. Meanwhile, parameterizing the dark-energy equation of state offers a superior compatibility to a wide variety of scalar fields~\cite{35de2008calibrating}. Among others, the so-called vacuum metamorphosis (VM) model~\cite{13parker2000new,14parker2004acceleration} has shown an improved fit to the observational data~\cite{36di2018vacuum}. Despite the fiducial foundation, here we skip the reconstruction of the scalar field but focus on the properties of the equation of state $w$. In the original VM model, the dark energy evolves rapidly in fairly recent redshifts between $z=1.3$ and $z=0.7$. We replicate such behavior in the equation of state and mimic the evolution of the VM model in the context of our numerical scheme in Fig.~\ref{fig:phantom} where $q$ denotes the deceleration parameter.
The initial conditions are established as mentioned previously. The result shows a surprising degree of accuracy on the CMB acoustic scale (at the 0.2\% level) for the model with $\left<w\right> = -1.26$. In particular, we find that the tension between the measurement on the Hubble constant in opposite epochs is reconciled. Nevertheless, we have avoided the background assumption on the actual dark-energy component in which the need of incorporating the phantom is rational but lack of motivation.  This is an important issue to be discussed in the last section. 

\begin{figure}
	\scalebox{0.7}
	{\includegraphics{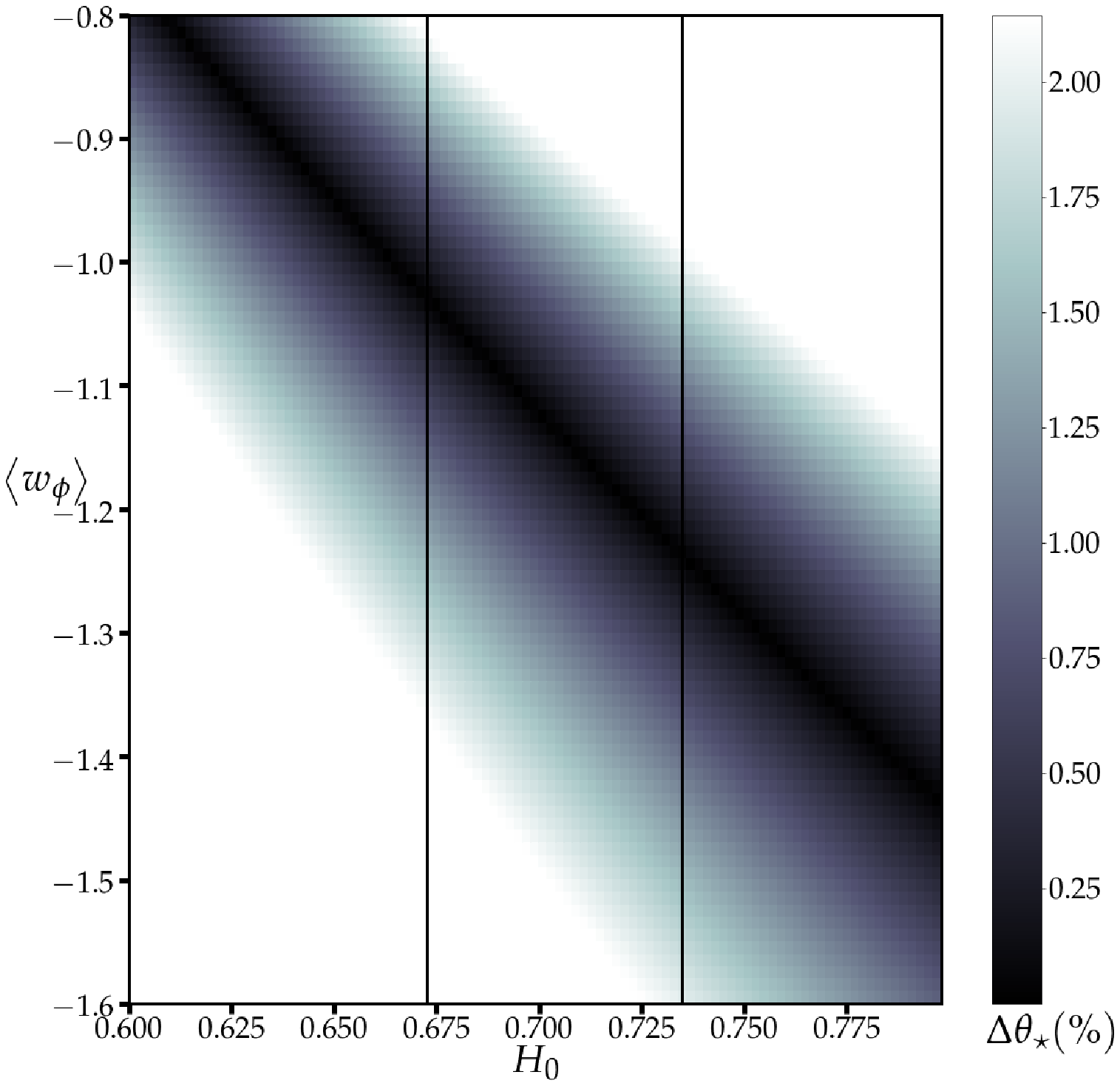}}
	\caption{\label{fig:H_0_w_test_final} The time-average $\left<w\right>$ vs. $H_0$ with the error on the angular size of the acoustic horizon. The vertical lines in the diagram indicate the representative values of the Hubble constant obtained respectively from the CMB anisotropy (left) and the local distance measurement on supernovae (right).}
\end{figure}

Though the VM model provides great conformity with the CMB acoustic scale, we are more interested in the reason why the involvement of a phantom is necessary. We thus plot the relation between the time averaged $\left<w\right>$ versus the Hubble constant $H_0$ in Fig.~\ref{fig:H_0_w_test_final} where the shade indicates the degree of accuracy on the CMB acoustic scale.
The averaged dark energy density around the last scattering surface $\Omega_{\phi,\star}$ has been ignored due to the rapidly evolving equation of state. The approximation on $\eta_\star$ is reduced to an equation dominated by $H_0$ thus affecting the values of energy density distributed to each component. Note that in some quintessence models (the part in Fig.~\ref{fig:H_0_w_test_final} where the time-average $\left<w\right>>-1$) $\Omega_{\phi,\star}$ may not be negligible, which in turn becomes a significant factor in the determination of the acoustic scale~\cite{25GQ}.  However, the information from Table I enables us to ignoring $\Omega_{\phi,\star}$ completely in the fitting test. 
\begin{table}
	\caption{\label{tab:table1}The result on three typical categories of dark energy under numerical calculations}
\begin{ruledtabular}
	\begin{tabular}{cccccccc}
		&GQ\footnotemark[1]&GQ&$\Lambda CDM$&VM\footnotemark[4]&VM\\
		&+Planck\footnotemark[2]&+R18\footnotemark[3]&&+Planck&+R18\\
		\hline
		$\Delta\theta_\star$& 20.85\% & 23.42\% & 0.013\% & 1.44\% & 0.049\% \\
		$\left<w\right>$& -0.902 & -0.898 &-1.0 & -1.261 & -1.290 \\
		$\Omega_{\phi,\star}$& 0.294 & 0.348 & $1.27\times10^{-9}$ & $2.08\times10^{-10}$ & $2.66\times10^{-10}$ \\
	\end{tabular}
\end{ruledtabular}
\footnotetext[1]{the quintessence model~\cite{25GQ}.}
\footnotetext[2]{the Planck collaboration data~\cite{12ade2016planck}.}
\footnotetext[3]{the distance measurement data~\cite{37riess20162}.}
\footnotetext[4]{the phantom model~\cite{36di2018vacuum}.}
\end{table}
Furthermore, the darkest region in the middle of Fig.~\ref{fig:H_0_w_test_final} reveals the acoustic scale with a minimal error, in which two substantial implications follow. First of all, the VM model seems sensible to explain the late time acceleration if we take the local measurement data into account. As a matter of fact, any phantom model can play such role provided that the associated time-average $\left<w\right>$ admits an appropriate value. Second, even if we accept the Hubble constant derived from CMB anisotropies, the equation of state will presume a value slightly less than $-1$ to match the result given by the Planck collaboration. Consequently, it is essential to comprise the phantom as the working hypothesis of a dynamical dark energy. 

\begin{figure}
	\scalebox{0.7}
	{\includegraphics{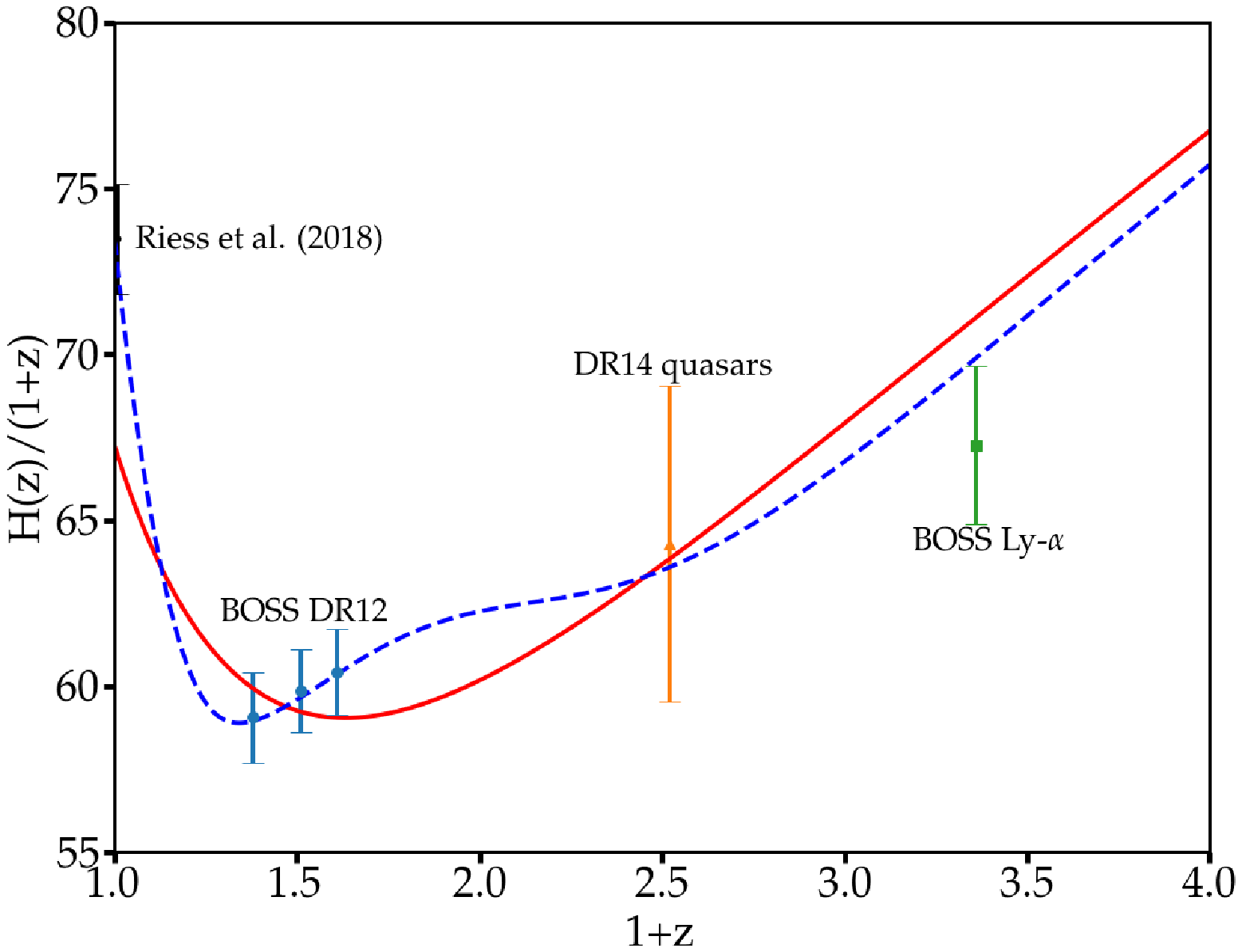}}
	\caption{\label{fig:H_0_plot_final} BAO measurement on $H(z)$ in different redshifts. The red line represents the Hubble parameter in the $\Lambda$CDM. The blue dash line describes the behavior of the dynamical dark energy model under consideration. }
\end{figure}

\begin{figure}
	\scalebox{0.7}
	{\includegraphics{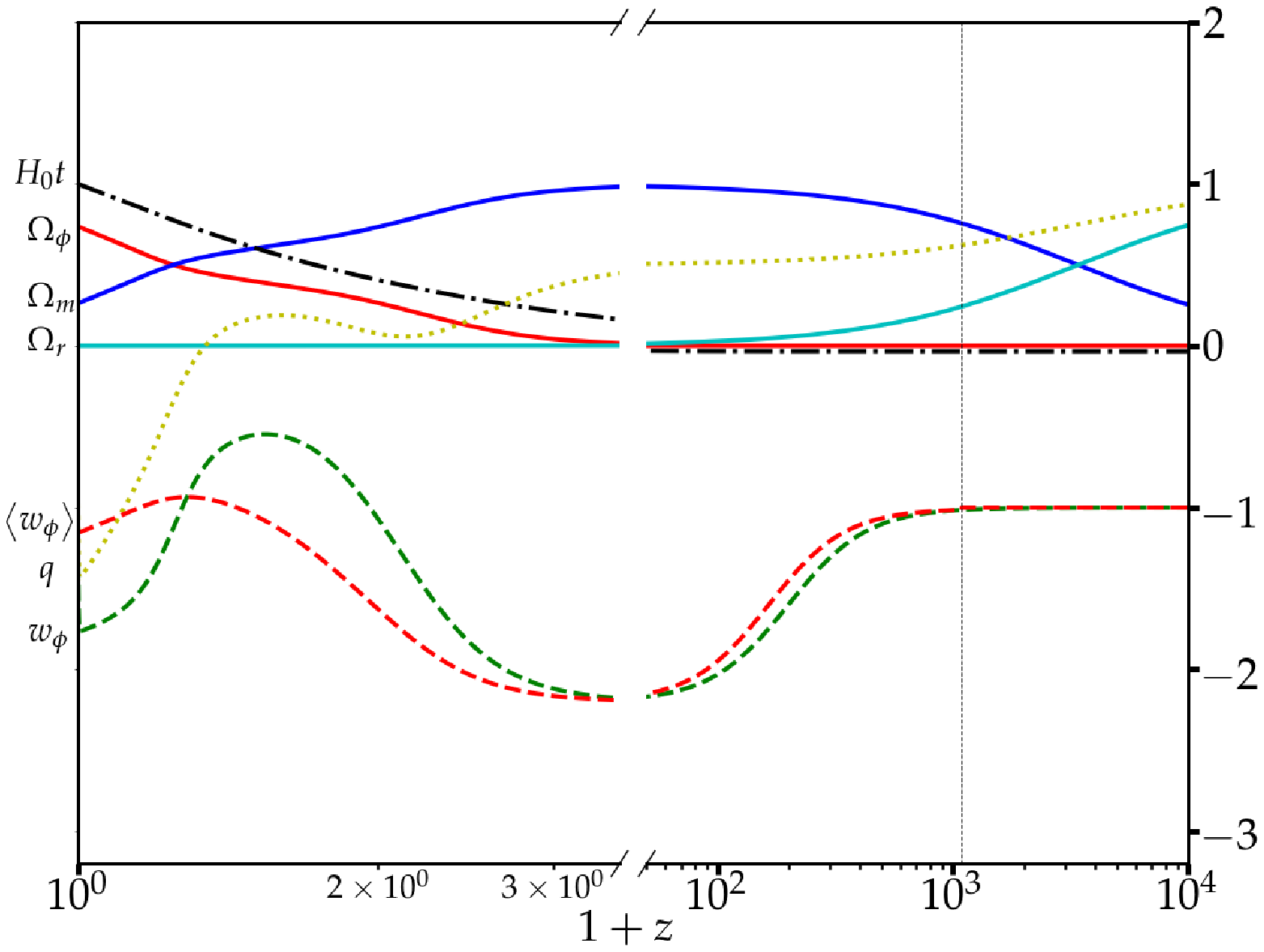}}
	\caption{\label{fig:BAO_test_final} Evolutions of cosmological parameters in the dynamical dark energy model with a $1.42\%$ error on $\theta_\star$ while maintaining $\left<w\right>=-1.15$. }
\end{figure}

We now add the consistency check on the measurements of baryon acoustic oscillations into our analysis. The adjustable equation of state allows for easy setups for fitting the dynamical dark energy with the recent data. The results are illuminated in Fig~\ref{fig:H_0_plot_final} and Fig~\ref{fig:BAO_test_final}.%
We find that by manipulating the equation of state, the Hubble parameter matches the released data on BAO competently. With the help from the formula,
\begin{equation}
\chi^2 = \sum_i\left( \frac{H(z_i)_{\rm obs}-H(z_i)_{\rm exp}}{H(z_i)_{\rm err}}\right),
\end{equation}
where ``obs",``exp" and ``err" denote the observational, the experimental and the error bar respectively, we perform the Chi-square test on the Hubble parameter. It shows that $\chi^2 = 5.85$ for the $\Lambda$CDM is reduced to $\chi^2 = 4.69$ for a dynamical dark energy. Moreover, the error on the acoustic peak still maintains at a low level as $\Delta\theta_\star = 1.42\%$. 

\begin{figure}
	\scalebox{0.7}
	{\includegraphics{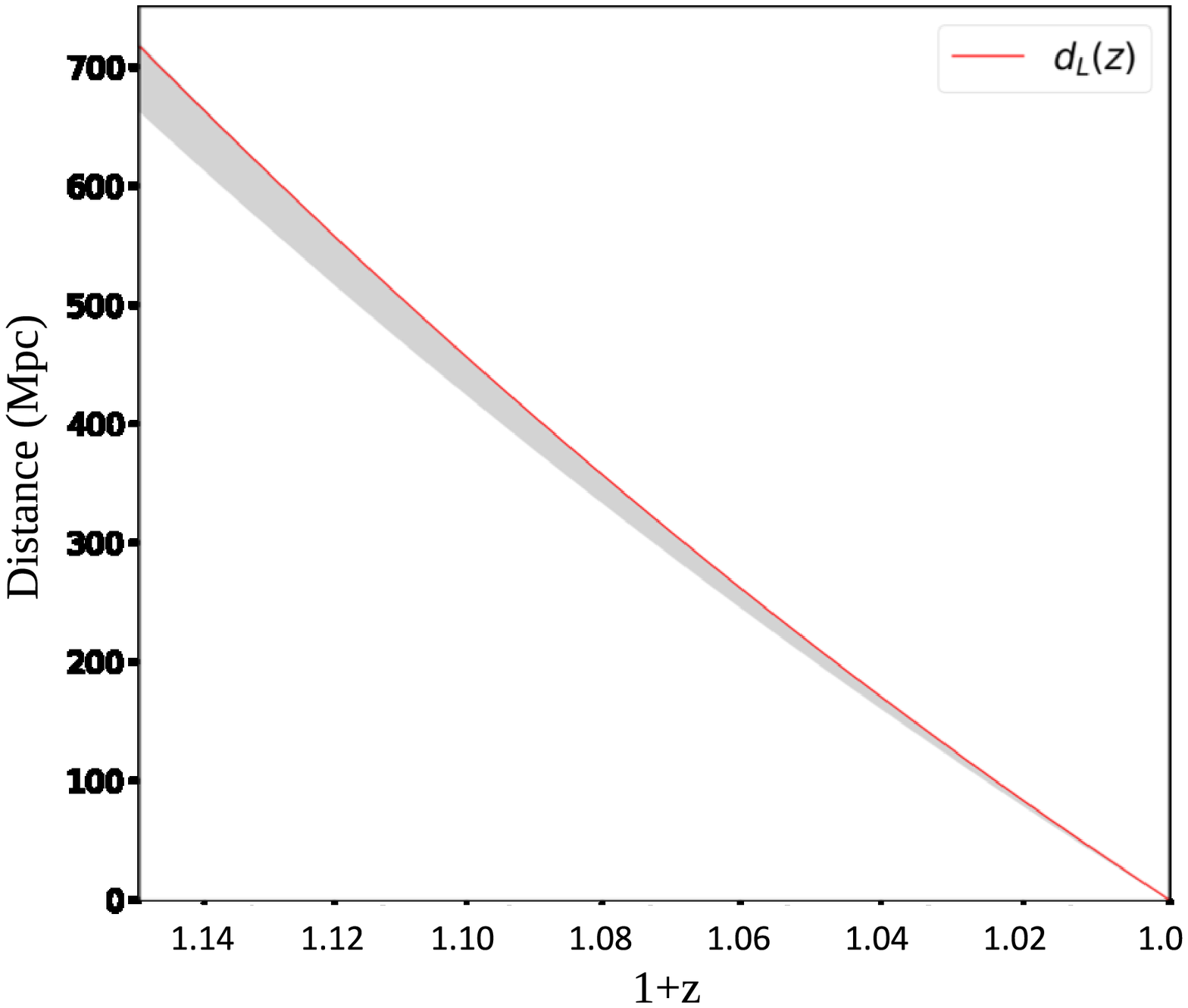}}
	\caption{\label{fig:d_l_z_relation} The luminosity distance-redshift relation in the low redshift region. The gray area specifies the 1-$\sigma$ error bar of $d_L(z)$, derived from the measurement of $H_0$ and $q_0$ in~\cite{camarena2020local}. The red solid line represents the luminosity distance of the model in Fig~\ref{fig:BAO_test_final}.}
\end{figure}

In order to match different local distance measurements, we check up the luminosity distance-redshift relation in the low redshifts according to 
\begin{equation}
d_L(z) = \frac{cz}{H_0}\left[ 1+\frac{(1-q_0)z}{2}+O(z^2)\right], 
\end{equation}
which depends only on two parameters, $H_0$ and $q_0$. Applying the data in~\cite{camarena2020local} where $H_0 = 75.35\pm1.68$ and $q_0 = -1.08\pm0.29$, we plot $d_L$ in the low redshift range up to $z\leq 0.15$ as in Fig~\ref{fig:d_l_z_relation}. Apparently, our model in Fig~\ref{fig:BAO_test_final} does locate in the area within $1\sigma$ error. 
  
In the meanwhile we also check on $\sigma_8$, the amplitude of density fluctuations on the scale of $8h^{-1}$Mpc, which is influenced by the matter energy density at the present time and the property of dark energy. It can be constrained by the model-independent measurement on the x-ray emission from galactic clusters. Mimicking the VM model in our first calculation, we argue that $\sigma_8$ under the dynamical dark energy assumes a similar value~\cite{36di2018vacuum}. Due to the phantomlike characteristics of $\langle w\rangle < -1$, the matter density $\Omega_m$ evolves in the same fashion in low redshifts, as illustrated in Fig.~\ref{fig:example}, ~\ref{fig:phantom} and ~\ref{fig:BAO_test_final}.  Consequently, the amplitude of density fluctuations in low redshifts among models under consideration should not deviate significantly, and the evolution of energy densities maintains much the same tendency of variation.

\section{\label{sec:level4}Conclusion}

Employing a generalized phenomenological approach, we examine the consistency in the evolution of cosmic backgrounds under the dynamical dark energy according to the acoustic scale $l_A$, the Hubble parameters $H(z)$, and the BAO measurements. Our numerical calculation shows that the error on the acoustic scale has reduced to $0.049\%$ in the target model, the vacuum metamorphosis, which highly supports the indispensable involvement of a phantom provided that we trust the local distance measurement on the Hubble constant $H_0$. In addition, the combined constraints imply that the dark energy may be more deeply phantom. The seemingly entanglement of the phantom with ordinary matters is further confirmed by the $\left<w\right>$-$H_0$ diagram, i.e. Fig.~\ref{fig:H_0_w_test_final} in our analysis. Moreover, the $H_0$ tension is arguably alleviated by the dynamics of the evolving dark energy. 

On the other hand, the trace of the phenomenologically reconstructed dark-energy equation of state seems to be complicated and almost beyond imagination, especially when taking the BAO measurement into account. Similarly, it may become too artificial to fit the hypothetical potential field of dark energy with all observational data. As the result, the phantom component might be a misleading issue whose authenticity is hard to distinguish. Perhaps predicting the dark energy by the equation of state but not the property of spacetime is inappropriate. This also brings up the coincident problem why the missing energy density evolves so rapidly to two-third in the present time. Apparently we still rely on the next generation of observations, e.g. the gravitational wave, to actually resolve the tension on the Hubble constant, and hopefully offer new perspectives on dark energy. We will need more fundamental theories to interpret the essence of the missing energy.  

\acknowledgments

This work is supported in part by the Ministry of Science and Technology of Taiwan, R.O.C. under Grants No. MOST 109-2112-M-001-003 (K.-W. Ng) and MOST 108-2112-M-003-001- (W. Lee) . 

\appendix

\section{\label{app:etaapp} the approximation to pin down the acoustic scale}
The analytical formulas for $\eta_{\star}$ and $\eta_0$ can be obtained as following: Consider the Friedmann equation with the reduced Planck mass $M_p^2\equiv(8\pi G)^{-1}$ in a spatially flat universe such that
\begin{equation}
H^2 = \frac{1}{3M_p^2}\left(\rho_m+\rho_r+\rho_\phi\right),
\end{equation}
where $\rho_\phi$ represents the energy density of the dark energy field. It can then be expressed in the form as
\begin{equation}
3M_p^2H^2\left[1-\Omega_\phi(t)\right] = 3M_p^2H_0^2\left(\Omega_{m,0}a^{-3}+\Omega_{r,0}a^{-4}\right).
\end{equation}
\subsection{\label{app:subsec1}$\eta_\star$ approximation}
Substituting $\Omega_\phi(t)$ with the constant average $\Omega_{\phi,\star}$ for the period around the last scattering surface, we have
\begin{equation}
\left(\frac{H}{H_0}\right)^2 = \frac{\Omega_{m,0}a^{-3}+\Omega_{r,0}a^{-4}}{1-\Omega_{\phi,\star}}.
\end{equation}
Converting the time coordinate to the conformal time $\eta$, the rate of the cosmic expansion can be written as
\begin{equation}
\left(\frac{da}{d\eta}\right)^2 = \frac{\Omega_{m,0}a+\Omega_{r,0}}{1-\Omega_{\phi,\star}}. 
\label{eq:a4}
\end{equation}
Thus, the conformal time at the last scattering surface can be obtained by integrating (\ref{eq:a4}) as
\begin{equation}
\eta_\star \simeq \frac{2\sqrt{1-\Omega_{\phi,\star}}}{\sqrt{\Omega_{m,0}}}\left(\sqrt{a_{\star}+\frac{\Omega_{r,0}}{\Omega_{m,0}}}-\sqrt{\frac{\Omega_{r,0}}{\Omega_{m,0}}}\,\right). \label{app1}
\end{equation}
\subsection{\label{app:subsec2}$\eta_0$ approximation}
With a constant equation of state $\left<w_\phi\right>$, the scaling law for the dark energy component can be characterized as $\Omega_\phi(t)\simeq\Omega_{\phi,0}a^{-3(1+\left<w_\phi\right>)}$. Therefore, the Friedmann equation in terms of the conformal time becomes
\begin{equation}
\left(\frac{da}{d\eta}\right)^2 = H_0^2\left[\Omega_{m,0}a+\Omega_{\phi,0}a^{(1-3\left<w_\phi\right>)}+\Omega_{r,0}\right].
\end{equation}
As the result, the conformal time at the present can be obtained as an integral as 
\begin{equation}
\eta_0 \simeq\frac{1}{H_0\sqrt{\Omega_{m,0}}}\int_{0}^{1}\left\{a+\frac{\Omega_{\phi,0}}{\Omega_{m,0}}a^{(1-3\left<w_\phi\right>)}+\frac{\Omega_{r,0}}{\Omega_{m,0}}\right\}^{-1/2}da \label{app2}
\end{equation}
\nocite{*}


\bibliography{yph01}

\end{document}